\documentstyle[11pt]{article}

\begin{document}

\begin{flushright}
{BIHEP-TH-95-30\\
hepph/9604423}
\end{flushright}
\begin{center}

{\LARGE 
The $x$-dependent helicity distributions for valence quarks in nucleons}

\vspace{5mm}

\renewcommand{\thefootnote}{\fnsymbol{footnote}}
{\large Bo-Qiang Ma} 

{\small 
CCAST(World Laboratory), P.O.Box 8730, Beijing 100080, and 
Institute of High Energy Physics, Academia Sinica, P.O.Box 918(4), 
Beijing 100039, China\footnote{Mailing address}

}
\vspace{5mm}

{\large \bf Abstract } \\
\end{center}

We derive simple relations between the polarized and unpolarized
valence quark distributions in a light-cone SU(6) quark-spectator 
model for nucleons. The explicit x-dependent
Wigner rotation effect for the light-flavor quarks is calculated. 
It is shown that the mass difference between the scalar
and vector spectators could reproduce the up and down valence
quark asymmetry that
accounts for the observed ratio $F_2^{n}/F_2^{p}$.
The proton, neutron, and deuteron polarization asymmetries, $A_1^p$, 
$A_1^n$, and $A_1^d$, are in agreement 
with the available data by taking into account the Wigner rotation
effect. The calculated $x$-dependent helicity distributions 
of the up and down valence quarks are compared with
the recent 
results from semi-inclusive hadron
asymmetries in polarized deep inelastic scattering by the Spin Muon
Collaboration. 
      
\vspace{10mm}
\break

The observation of the Ellis-Jaffe sum rule (EJSR) violation in
the inclusive polarized deep inelastic scattering experiments
at CERN \cite{EMC,SMC} and SLAC \cite{E142,E143} 
has received an extensive attention
on the spin content of nucleons. The experimental data
of the integrated spin-dependent structure functions for the nucleons
are generally
understood to imply that the sum of the up, down, and strange
quark helicities in the nucleon is much smaller than the nucleon
spin \cite{Ma93}. 
There has been a number of possible interpretations \cite{Ans94}
for the EJSR violation, and the
quark helicity distributions for 
each flavor 
are quite different
in these interpretations. It is clear that precise 
experimental measurements about the explicit 
quark helicity distributions for each flavor
will be of crucial importance to test various interpretations of the     
EJSR violation.  

The semi-inclusive processes in polarized deep inelastic 
scattering have been proposed to provide
further informations 
about the helicity distributions for different quark flavors
\cite{Fra89,Clo91,Lu95,Ell95}.  
The measurements of 
the helicity distributions 
for the valence up (u) and down (d) quarks from
semi-inclusive hadron charge asymmetries in polarized deep inelastic
scattering 
have been taken by the Spin Muon Collaboration (SMC) recently
\cite{NSMC,NSMCN}. 
The
progress in experiments 
requires improved theoretical calculations
of each flavor valence and sea quark helicity distributions 
that can be measured in future experiments.
There have been attempts to calculate the quark helicity
distributions for each flavor from general 
QCD arguments in light-cone formalism
\cite{Bro95}, by assuming simple relations with the unpolarized
quark distributions \cite{Bou95}, or from parametrizations
\cite{Geh95}.
The purpose of this paper is to investigate the 
quark helicity distributions for the valence u and d quarks
in a light-cone SU(6) quark-spectator 
model for nucleons. 
We derive simple relations between 
polarized and unpolarized
valence quark distributions
by taking into account the explicit flavor asymmetry
due to the difference between the 
scalar and vector spectators and the Wigner rotation
effect due to the internal quark transversal motions
\cite{Ma93,Ma91,Bro94,Web92}.

The deep inelastic lepton nucleon scattering is well
described by the impulse approximation picture of 
the quark-parton model \cite{parton,Bro82,Ma86} 
in which the incident lepton scatters incoherently off a parton
in the nucleon 
with the remaining nucleon constituents treated as
a quasi-particle spectator. 
From this picture one
can calculate the valence quark distributions  
in the quark-diquark model where the single valence quark
is the scattered parton and 
the non-interacting diquark serves to provide the quantum number
of the spectator
\cite{QM,Car75,Sch88}. 
The proton wave function in the quark-diquark model \cite{Pav76}
is written as 
\begin{equation}
\Psi _p^{\uparrow, \downarrow}(qD)
=sin\theta\, \varphi_{V}\, 
|qV>^{\uparrow, \downarrow} 
+ cos\theta\,
\varphi_{S}\, |qS>^{\uparrow, \downarrow},
\label{eq:pwf}
\end{equation}
with 
\begin{equation}
\begin{array}{clcr}

|qV>^{\uparrow, \downarrow}
=\pm \frac{1}{3}[V_0(ud)u^{\uparrow, \downarrow}-
\sqrt{2}V_{\pm 1}(ud)u^{\downarrow, \uparrow}
-\sqrt{2}V_0(uu)d^{\uparrow, \downarrow}
+2V_{\pm 1}(uu)d^{\downarrow, \uparrow}]; \\

|qS>^{\uparrow, \downarrow}
=S(ud)u^{\uparrow, \downarrow},
\label{eq:pwfVS}
\end{array}
\end{equation}
where $V_{s_z}(q_{1}q_{2})$ stays for a $q_1 q_2$ vector diquark 
Fock state with third spin 
component $s_z$, 
$S(ud)$ stays for a $ud$ scalar diquark Fock state,
$\varphi_{D}$
stays for the momentum space wave function of the 
quark-diquark with
$D$ representing 
the vector ($V$) or scalar ($S$) diquarks, 
and $\theta$ is a mixing angle that
breaks SU(6) symmetry at $\theta\neq \pi /4$. 
In this paper we choose the bulk SU(6) symmetry case $\theta=\pi /4$.
From Eq.~(\ref{eq:pwf}) we get the unpolarized quark distributions
\begin{equation}
\begin{array}{clcr}
u_{v}(x)=\frac{1}{2}a_S(x)+\frac{1}{6}a_V(x);\\
d_{v}(x)=\frac{1}{3}a_V(x),
\label{eq:ud}
\end{array}
\end{equation}
where $a_D(x)$ ($D=S$ or $V$) is normalized such
that $\int_0^1 d x a_D(x)=3$ and denotes the amplitude for the quark
$q$ is scattered while the spectator is in the diquark state $D$
\cite{Car75}.
Therefore we can write, 
by assuming the isospin symmetry between the proton and the neutron,
the unpolarized structure functions for nucleons, 
\begin{equation}
\begin{array}{clcr}
F^p_2(x)=x s(x)+\frac{2}{9} x a_S(x)+\frac{1}{9} x a_V(x);\\
F^n_2(x)=x s(x)+\frac{1}{18} x a_S(x)+\frac{1}{6} x a_V(x),
\label{eq:Fpn}
\end{array}
\end{equation}
where $s(x)$ denotes the contribution from the sea. 

Exact SU(6) symmetry provides the relation $a_S(x)=a_V(x)$ 
which implies the valence flavor symmetry $u_{v}(x)=2 d_{v}(x)$. This
gives the prediction $F^n_2(x)/F^p_2(x)\geq 2/3$ for
all $x$ \cite{Kut71} and is ruled out by the experimental
observation $F^n_2(x)/F^p_2(x) <  1/2$ for $x \to 1$.
It has been a well established fact that the valence flavor
symmetry $u_{v}(x)=2 d_{v}(x)$ does not hold and the explicit
$u_{v}(x)$ and $d_{v}(x)$ can be parameterized from  
the combined
experimental data from deep inelastic 
scatterings of electron (muon) and neutrino (anti-neutrino) 
on the proton and the neutron {\it et al.}.  
In this sense, any theoretical calculation of quark distributions 
should reflect the flavor asymmetry between 
the valence u and d quarks in a reasonable picture.
We will show in the following that the mass
difference between the scalar
and vector spectators can reproduce the up and down valence
quark asymmetry that 
accounts for the observed ratio $F_2^{n}(x)/F_2^{p}(x)$ at large $x$.

From the well-known quark-parton model \cite{parton} we know that 
it is proper to describe deep inelastic
scattering 
as the sum of incoherent scatterings
of the incident lepton on the partons 
in the infinite momentum frame or in the light-cone formalism
\cite{Bro82,Ma86}.
In the Bjorken limit $Q^2 \to \infty$ and $\nu \to \infty$
with $x=Q^2/2M\nu$ ranging from 0 to 1,
the amplitude for the quark
$q$ is scattered while the spectator in the spin state $D$
can be written as
\begin{equation}
a_D(x)
\propto \int [{\rm d}^2 {\bf k}_{\perp}] |\varphi_{D}(x,{\bf k}_{\perp})|^2.
\end{equation}
We adopt the Brodsky-Huang-Lepage prescription \cite{BHL} for the
light-cone momentum space wave function of the 
quark-spectator 
\begin{equation}
\varphi_{D}(x,{\bf k}_{\perp})
=A_{D} exp\{-\frac{1}{8\beta^2_{D}}[\frac{m^2_q+{\bf k}^2_{\perp}}{x}
+\frac{m^2_D+{\bf k}^2_{\perp}}{1-x}]\},
\label{eq:BHL}
\end{equation}
where ${\bf k}_{\perp}$ is the internal quark transversal momentum,
$m_q$ and $m_D$ are the masses for the quark $q$ and spectator $D$,
and $\beta_D$ is the harmonic oscillator scale parameter.
The values of the parameters $\beta_D$, $m_q$ and $m_D$ 
can be adjusted by fitting the hadron properties
such as the electromagnetic form factors, 
the mean charge radiuses, and the
weak decay constants {\it et al.} in the relativistic light-cone
quark model \cite{LCQM}. 
In this paper we simply adopt $m_q=330$~MeV and
$\beta_D=330$~MeV as in the scale often adopted in literature \cite{LCQM}. 
The masses of the scalar and vector spectators should
be different taking into account the spin force from color magnetism
or alternatively from instantons \cite{Web92}. 
We choose, e.g., $m_S=600$ MeV and $m_V=900$ MeV as estimated
to explain the N-$\Delta$ mass difference. The mass difference
between the scalar and vector spectators
causes difference between $a_S(x)$ and $a_V(x)$, 
and the flavor asymmetry between the valence quark distribution
functions $u_v(x)$ and $d_v(x)$ can thus manifest itself. 
In Fig.~1 we present the calculated ratio $F_2^n(x)/F_2^p(x)$ 
by neglecting the sea contribution $s(x)$. The calculated
results are in reasonable agreement 
with the experimental data \cite{F2np}, at least for large $x$, and this
supports the quark-spectator picture of deep inelastic scattering
in which the difference between the scalar and vector
spectators is important to reproduce the explicit
SU(6) symmetry breaking while the bulk SU(6) symmetry of the 
quark model still holds. The above result is in agreement 
with the bag model
result of the effect on the structure functions due to the mass difference
between the spectator scalar and vector diquarks \cite{Clo88}.  

We now turn our attention to the polarized quark distributions.
We should take into account the contribution from the
Wigner rotation which is one important ingredient in
the quark-parton model and light-cone descriptions of polarized
deep inelastic scattering \cite{Ma93,Ma91,Bro94,Web92}. 
The quantity $\Delta q$ measured in polarized deep inelastic
scattering is defined by the axial current matrix element
\begin{equation}
\Delta q=<p,\uparrow|\overline{q} \gamma^{+} \gamma_{5} q|p,\uparrow>.
\end{equation}
In the light-cone or quark-parton descriptions,
$\Delta q (x)=q^{\uparrow}(x)-q^{\downarrow}(x)$,
where $q^{\uparrow}(x)$ and $q^{\downarrow}(x)$ are the probability
of finding a quark or antiquark with longitudinal momentum
fraction $x$ and polarization parallel or antiparallel
to the proton helicity in the infinite momentum frame.
However, in the proton rest frame, one finds,
\begin{equation}
\Delta q (x)
=\int [{\rm d}^2{\bf k}_{\perp}] W_D(x,{\bf k}_{\perp}) 
[q_{s_z=\frac{1}{2}}
(x,{\bf k}_{\perp})-q_{s_z=-\frac{1}{2}}(x,{\bf k}_{\perp})],
\end{equation}  
with
\begin{equation}
W_D(x,{\bf k}_{\perp})=\frac{(k^+ +m)^2-{\bf k}^2_{\perp}}
{(k^+ +m)^2+{\bf k}^2_{\perp}} 
\end{equation}
being the contribution from the relativistic effect due to
the quark transversal motions, 
$q_{s_z=\frac{1}{2}}(x,{\bf k}_{\perp})$
and $q_{s_z=-\frac{1}{2}}(x,{\bf k}_{\perp})$ being the probability
of finding a quark and antiquark with rest mass $m$
and with spin parallel and anti-parallel to the rest proton
spin, and $k^+=x {\cal M}$ where 
${\cal M}=\frac{m^2_q+{\bf k}^2_{\perp}}{x}
+\frac{m^2_D+{\bf k}^2_{\perp}}{1-x}$.
The Wigner rotation factor $W_D(x,{\bf k}_{\perp})$ ranges
from 0 to 1; thus $\Delta q$ measured 
in polarized deep inelastic scattering cannot be  
identified
with the spin carried by each quark flavor in the proton
rest frame. This can be understood from the fact that the vector 
sum of the constituent spins for a composite system is not Lorentz
invariant by taking into account the relativistic 
effect from the Wigner rotation \cite{Ma93}. 

From Eq.~(\ref{eq:pwf}) we get the spin distribution probabilities
in the quark-diquark model
\begin{equation}
\begin{array}{clcr}
u_V^{\uparrow}=\frac{1}{18};\;\;
u_V^{\downarrow}=\frac{2}{18};\;\;
d_V^{\uparrow}=\frac{2}{18};\;\;
d_V^{\downarrow}=\frac{4}{18};\\
u_S^{\uparrow}=\frac{1}{2};\;\;
u_S^{\downarrow}=0;\;\;
d_S^{\uparrow}=0;\;\;
d_S^{\downarrow}=0.\\
\label{eq:udVS}
\end{array}
\end{equation}
From the above discussions about the Wigner rotation,
we can write the quark helicity distributions
for the u and d quarks 
\begin{equation}
\begin{array}{clcr}
\Delta u_{v}(x)=u_{v}^{\uparrow}(x)-u_{v}^{\downarrow}(x)=
-\frac{1}{18}a_V(x)W_V(x)+\frac{1}{2}a_S(x)W_S(x);\\
\Delta d_{v}(x)=d_{v}^{\uparrow}(x)-d_{v}^{\downarrow}(x)
=-\frac{1}{9}a_V(x)W_V(x),
\label{eq:sfdud}
\end{array}
\end{equation}
where $W_D(x)$ is the correction factor due the Wigner rotation.
From Eq.~(\ref{eq:ud}) one gets 
\begin{equation}
\begin{array}{clcr}
a_S(x)=2u_v(x)-d_v(x);\\
a_V(x)=3d_v(x).
\label{eq:qVS}
\end{array}
\end{equation}
Combining Eqs.~(\ref{eq:sfdud}) and (\ref{eq:qVS}) we have
\begin{equation}
\begin{array}{clcr}
\Delta u_{v}(x)
    =[u_v(x)-\frac{1}{2}d_v(x)]W_S(x)-\frac{1}{6}d_v(x)W_V(x);\\
\Delta d_{v}(x)=-\frac{1}{3}d_v(x)W_V(x).
\label{eq:dud}
\end{array}
\end{equation}
Thus we arrive at simple relations between the polarized
and unpolarized quark distributions for the valence u and d
quarks. We can calculate the quark helicity distributions
$\Delta u_{v}(x)$ and $\Delta d_{v}(x)$ from the unpolarized
quark distributions $u_{v}(x)$ and $d_{v}(x)$ by relations
(\ref{eq:dud}),
once the detailed $x$-dependent Wigner rotation
factor $W_D(x)$ is known. On the other hand, we can also
use relations (\ref{eq:dud}) to study 
$W_S(x)$ and $W_V(x)$, once there are good quark distributions 
$u_{v}(x)$, $d_{v}(x)$,
$\Delta u_{v}(x)$, and $\Delta d_{v}(x)$ from experiments.
From another point of view, the relations (\ref{eq:dud})
can be considered as the results of the conventional
SU(6) quark model \cite{QM,Car75,Sch88} 
by explicitly taking into account the Wigner rotation effect
and the flavor asymmetry introduced by the
mass difference between the scalar and vector
spectators, 
thus any evidence for the 
invalidity of Eq.~(\ref{eq:dud}) will be useful to reveal 
new physics beyond the SU(6) quark model.

Encouraged by our above calculation 
of the valence u and d flavor asymmetry that 
accounts for the observed ratio $F_2^{n}(x)/F_2^{p}(x)$,  
we calculate the $x$-dependent Wigner rotation factor
$W_D(x)$ in the light-cone SU(6) quark-spectator  
model by adopting the light-cone wave
function Eq.~(\ref{eq:BHL}). 
The calculated results are presented in Fig.~2 from which we
notice the slight asymmetry between $W_S(x)$ (full curve) and $W_V(x)$
(dotted curve).
Considering only the valence quark contributions, 
we  can write the 
spin-dependent structure functions $g_1^p(x)$ and $g_1^n(x)$
for the proton and the neutron by
\begin{equation}
\begin{array}{clcr}
g_1^p(x)=\frac{1}{2}[\frac{4}{9}\Delta u_v(x)
+\frac{1}{9}\Delta d_v(x)]
=\frac{1}{18}[(4 u_v(x)-2 d_v(x))W_S(x)-d_v(x)W_V(x)];\\
g_1^n(x)=\frac{1}{2}[\frac{1}{9}\Delta u_v(x)
+\frac{4}{9}\Delta d_v(x)]
=\frac{1}{36}[(2 u_v(x)-d_v(x))W_S(x)-3 d_v(x)W_V(x)].
\label{eq:g1pn}
\end{array}
\end{equation}
The proton, neutron, and deuteron polarization asymmetries, $A_1^p$,
$A_1^n$, and $A_1^d$, are directly measured in experiments and are
expressed by 
$A_1^N(x)=2 x g_1^N(x)/F_2^N(x)$, where $N$ denotes $p$, $n$, and
$d$.
We thus can adopt 
one set of existing unpolarized quark distribution parametrizations
\cite{GRV94} 
and the calculated $W_S(x)$ and $W_V(x)$ to calculate 
$A_1^N$.
We present in Fig.~3 the calculated polarization asymmetries
$A_1^N$ 
for the cases with asymmetric Wigner rotation ($W_V(x)\neq W_S(x)$) of
fig.2 (full curves), no Wigner rotation ($W_S=W_V=1$) (dotted curves), 
and large asymmetric Wigner rotation 
(dashed curves, see discussions
below) respectively.  
We see that the calculated $A_1^N$
with Wigner rotation 
are in agreement  
with the experimental data, at least for $x \geq 0.1$.
The large asymmetry between $W_S(x)$ and $W_V(x)$
has consequence for a better fit of the data.

As we consider only the valence quark contributions
to $g_1^p(x)$ and $g_1^n(x)$, we should not expect to fit
the Ellis-Jaffe sum data from experiments. This leaves
room for additional contributions from sea quarks
or other sources.
We point out, however, it is possible to reproduce
the observed Ellis-Jaffe sums $\Gamma_1^p=\int_0^1 g_1^p(x){\rm d} x$
and $\Gamma_1^n=\int_0^1 g_1^n(x){\rm d} x$ 
within the light-cone SU(6) quark-spectator model
by introducing a large asymmetry between the Wigner rotation
factors $W_S$ and $W_V$ for the scalar and vector spectators.
For example, we need $<W_S>=0.56$ and $<W_V>=0.92$
to produce $\Gamma_1^p=0.136$ and $\Gamma_1^n=-0.03$
as observed in experiments \cite{SMC,E143}. 
This can be achieved by adopting a large difference between 
$\beta_S$ and $\beta_V$ which should be adjusted by 
fitting other nucleon properties in the model \cite{Web92}.
The results of 
an example calculation of such kinds 
$A_1^p(x)$, $A_1^n(x)$, and $A_1^d(x)$ are
plotted (dashed curves) 
in Fig.~3 and the agreement with the data is good. 
This may suggest that the explicit
SU(6) asymmetry could be also used to explain the EJSR
violation (or partially) 
within a bulk SU(6) symmetry scheme of the quark model, or
we take this as a hint for other SU(6) breaking source in additional to
the SU(6) quark model.
Of course, the magnitude of the SU(6) asymmetry or breaking 
should be constrained
and estimated
from considerations of other nucleon properties.  

The validity of the results in this paper can be tested from
experimental measurements of $F_2^n(x)/F_2^p(x)$, 
$d(x)/u(x)$, 
and $\Delta d(x)$
near $x \to 1$. In this paper the dominant contribution
to polarized and unpolarized quark distributions at large $x$ is
from $u_v^{\uparrow}(x)$ due to the suppression of $a_V(x)$
in comparison with $a_S(x)$. This gives the prediction 
$F_2^n(x)/F_2^p(x)=1/4$ for $x \to 1$ and is consistent
with most parametrizations of unpolarized quark distributions.
However, there is an alternative perturbative QCD expectation
that the contribution
from the u and d valence quarks are 5:1 at large $x$ and this gives
the prediction $F_2^n(x)/F_2^p(x)=3/7$ for $x \to 1$ \cite{Bro95,Far75}.  
It has been recently suggested \cite{Mel96} 
that the measured ratio $F_2^n(x)/F_2^p(x)$ is
consistent with this expectation after taking into account Fermi
motion, binding and nucleon off-shell effects in the deuteron. 
This will be of crucial importance for later parametrizations
of quark distributions, once this conclusion is confirmed. 
The ratio of neutrino and anti-neutrino cross sections
on protons can provide further information about the
ratio $d(x)/u(x)$ near $x=1$ and test the two predictions
of the quark model ($d(x)/u(x) \to 0$) and the perturbative
QCD ($d(x)/u(x) \to 1/5$) \cite{Far77}, and this can avoid
the uncertainties in the ratio $F_2^n(x)/F_2^p(x)$
due to the nuclear effect. 
However, the available experimental data \cite{neutrino}
are compatible with the two
predictions and more precise data are needed for a clear
distinction of the two cases.  

We see from Eq.~(\ref{eq:dud}) that $\Delta d(x) \leq 0$
in this paper, and this is in consistent with some other results
\cite{Bou95,Geh95} and the quark model prediction
$\Delta d(x)/d(x) \to -1/3$ near $x \to 1$
\cite{QM,Car75,Sch88,Mel96}. 
The measurements of 
the explicit $x$-dependent quark 
helicity distributions 
for the valence u and d quarks have been taken
by the Spin Muon Collaboration from
semi-inclusive hadron charge asymmetries in polarized deep inelastic
scattering
\cite{NSMC,NSMCN}. 
This will put constraints on the theoretical calculations
of the quark helicity distributions for each flavor. 
In Fig.~4 we present
our calculated quark helicity distributions 
$\Delta u_{v}(x)$ and $\Delta d_{v}(x)$ and compare them with the
recent SMC data.
The data are still not precise enough for making detailed comparison,
but the agreement with $\Delta u_{v}(x)$ seems to be good.
It seems that the agreement with $\Delta d_{v}(x)$ is poor
and there is somewhat evidence for additional source of negative
helicity contribution to the valence d quark beyond the 
conventional quark model. Some progress have been made in this direction
and will be given elsewhere \cite{Bro96}.
As in the case of the ratio $F_2^n(x)/F_2^p(x)$ near $x \to 1$,
the quark model prediction $\Delta d(x)/d(x) \to -1/3$ is also
different from the perturbative QCD prediction 
$\Delta d(x)/d(x) \to 1$ at large
$x$ \cite{Bro95} and the available SMC data are still not possible
to make a clear distinction between the two predictions.
 
Similar to the $x \to 1$ behaviors, the $x \to 0$ behaviors
for the bulk SU(6) quark model results in this
paper and the perturbative QCD expectation \cite{Bro95} are also
different. In the quark model consideration of
this paper, the valence quarks provide the quantum
number of the nucleons, thus it is difficult to expect 
$q_v^{\uparrow}(x) = q_v^{\downarrow}(x)$ for the valence quarks
at $x=0$ without introducing large bulk SU(6) breaking mechanism. 
However, in the perturbative QCD consideration
one expects $q^{\uparrow}(x)-q^{\downarrow}(x) \to 0$ near $x \to 0$
corresponding to the fact that long range correlations disappear for
large rapidity separation and 
the unpolarized and polarized valence quark distributions
have different Regge behavior at $x=0$.
Therefore the $x \to 0$ behaviors
of the helicity distributions for the valence u and d quarks are also
important to distinguish between the quark model and perturbative
QCD predictions,
as in the case of the $x \to 1$ behaviors of
$F_2^n(x)/F_2^p(x)$, $d(x)/u(x)$ and $\Delta d(x)/d(x)$.   
From above discussions, improved precision in experiments 
in the $x \to 1$ and $x \to 0$ end-point regions 
are highly needed for extracting    
clear information of the quark flavor and helicity distributions
and revealing more about the explicit quark structure of nucleons.

In summary, we derived in this paper simple relations
between the polarized and unpolarized quark distributions
for the valence up and down quarks 
in a light-cone SU(6) quark-spectator 
model for nucleons. The explicit $x$-dependent
Wigner rotation effect for the light-flavor quarks is calculated. 
It is shown that the mass difference between the scalar
and vector spectators could reproduce the up and down valence
quark asymmetry that 
accounts for the observed ratio $F_2^{n}/F_2^{p}$.
The calculated proton, neutron, and deuteron polarization asymmetries, 
$A_1^p$, $A_1^n$, and $A_1^d$, 
are in agreement  
with the available data by taking into account the Wigner rotation
effect. The calculated $x$-dependent helicity distributions 
of the up and down valence quarks are compared with
the recent results from semi-inclusive hadron
asymmetries in polarized deep inelastic scattering by the Spin Muon
Collaboration. 

\noindent 
{\bf Acknowledgment}

The author would like to thank many useful discussions 
with S.~J.~Brodsky, H.~J.~Weber,
A.~Sch\"afer, and Z.~-t.~Liang 
during this work.  He is also grateful to W.~Wi\'slicki for
providing information of the SMC results.
This work was supported in part by the Chinese National Science
Foundation under grant no.19445004.

\break

\break
\noindent
{\large \bf Figure Captions}
\renewcommand{\theenumi}{\ Fig.~\arabic{enumi}}
\begin{enumerate}
\item
The ratio $F_2^n/F_2^p$ as a function of the Bjorken scaling variable
$x$. The curve is the calculated result in this work
and the data ($\bullet$) are from the revised NMC measurement \cite{F2np}.
\item
The $x$-dependent Wigner rotation factor $W_D(x)$ 
in the light-cone SU(6) quark-spectator model.
The full and dotted curves are the calculated $W_S(x)$ and $W_V(x)$
in this work. 
\item
The spin asymmetries $A_1^N(x)$ (a) for the proton,
(b) for the neutron, and (c) for the deuteron
as functions
of the Bjorken scaling variable $x$. 
The curves are the calculated 
$A_1^N(x)$ in the light-cone SU(6) quark-spectator model 
with the Gl\"uck-Reya-Vogt (GRV)
LO set of the unpolarized quark distribution parametrizations.
The full curves are the results for slight asymmetric
Wigner rotation factors $W_S(x)$ and $W_V(x)$ of fig.2; the dotted
curves are the results without Wigner rotation ($W_V=W_S=1$);
and the dashed curves are the results for a large asymmetry 
between $W_S(x)$ and $W_V(x)$ 
by adopting $\beta_S=500$ MeV and $\beta_V=200$ MeV.
The data
are from EMC($\bigtriangleup$), SMC($\Box$), 
E142($\Diamond$), and E143($\bigcirc$) experiments
\cite{EMC,SMC,E142,E143}.
\item
The quark helicity distributions (a) $x \Delta u_{v}$
and (b) $x \Delta d_v$ as functions of the Bjorken scaling
variable $x$. The thick full, dotted, and dashed curves
are the calculated results for slight asymmetric Wigner
rotation, no Wigner rotation, and large asymmetric
Wigner rotation respectively corresponding to fig.3. 
The thin full curves
represent limits given by unpolarized quark distributions
(i.e., $|\Delta q_v| \leq q_v$). The data ($\bullet$) are 
the SMC results from semi-inclusive hadron
asymmetries in polarized deep inelastic scattering \cite{NSMCN}.

\end{enumerate}

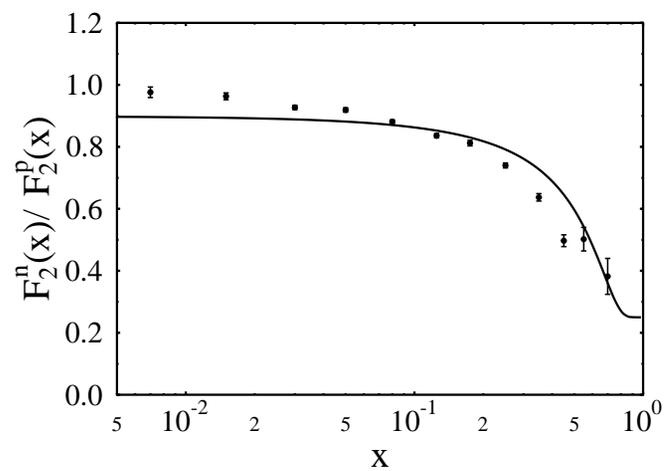
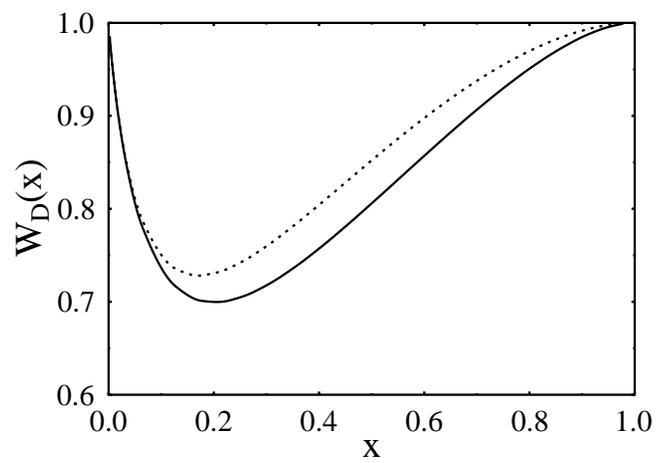
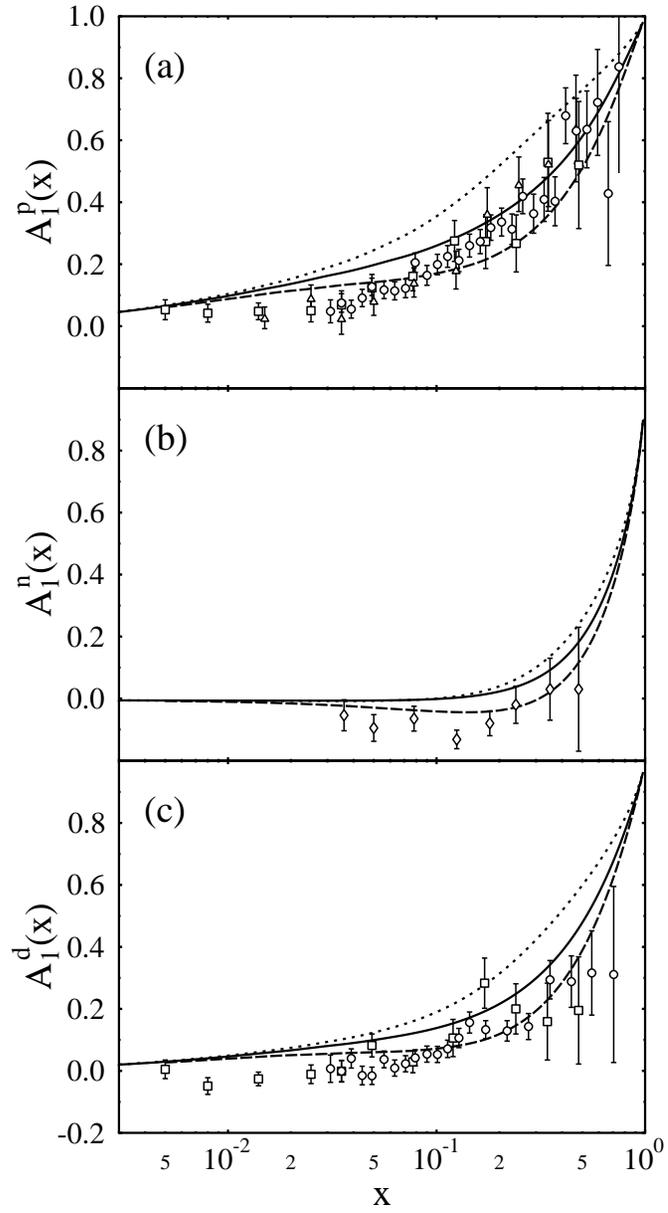
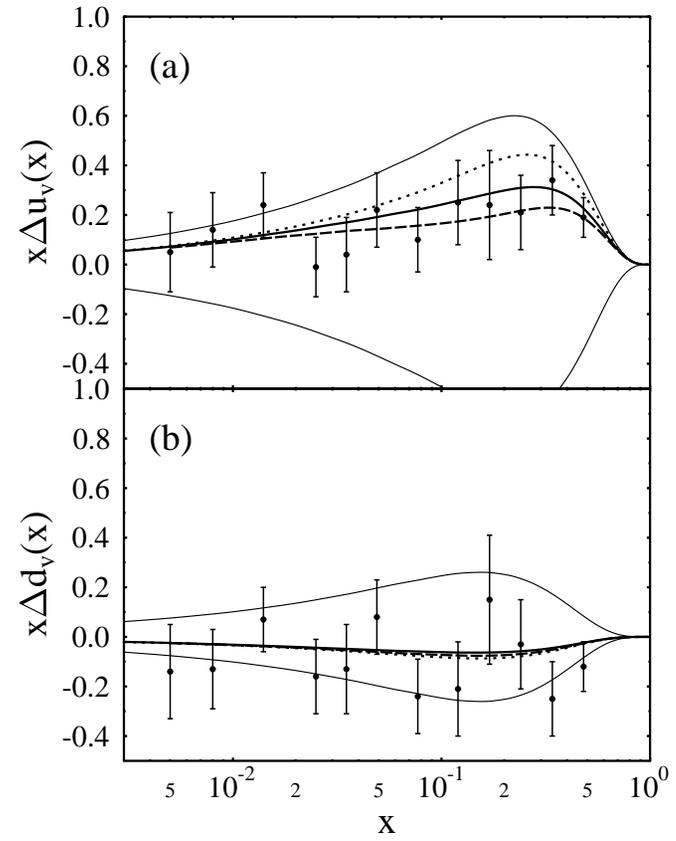
\end{document}